\documentclass{article}
\usepackage{spconf,amsmath,amssymb,epsfig}

\title{The Convergence of Machine Learning and Communications}
\name{Wojciech Samek$^1$, Slawomir Stanczak$^{1,2}$, Thomas Wiegand$^{1,2}$}%\thanks{This work was supported by the German Ministry for Education and Research as Berlin Big Data Center BBDC (01IS14013A).}}
\address{$^1$Fraunhofer Heinrich Hertz Institute, 10587 Berlin, Germany\\
$^2$Technische Universit\"at Berlin, 10587 Berlin, Germany}
\begin{document}
%\ninept
%
\maketitle
\begin{abstract} \em
The areas of machine learning and communication technology are converging.
Today's communications systems generate a huge amount of traffic data, which can help to significantly enhance the design and management of networks and communication components when combined with advanced machine learning methods. Furthermore, recently developed end-to-end training procedures offer new ways to jointly optimize the components of a communication system.
Also in many emerging application fields of communication technology, e.g., smart cities or internet of things, machine learning methods are of central importance.
This paper gives an overview over the use of machine learning in different areas of communications and discusses two exemplar applications in wireless networking. 
Furthermore, it identifies promising future research topics and discusses their potential impact.
\end{abstract}

%%%%%%%%%%%%%%%%%%%%%%%%%%%%%%%%%%%%%%%%%%%%%%%%%%%%%%%%%%%%%%%
\section{Introduction}
\label{sec:intro}
We are currently observing a paradigm shift towards ``smart'' communication networks that take advantage of network data. In fact, modern communication networks, and in particular mobile networks, generate a huge amount of data at the network infrastructure level and at the user/customer level. The data in the network contain a wealth of useful information such as location information, mobility and call patterns. The vision of network operators is to either enable new businesses through the provisioning of this data (or the information contained in it) to external service providers and customers or to exploit the network data for in-house services such as network optimization and management.

In order to make the vision reality, there is a strong need for the development and implementation of new machine learning methods for big data analytics in communication networks. The objective of these methods is to extract useful information from the network data while taking into account limited communication resources, and then to leverage this information for external or in-house services.

Moreover, machine learning methods are core part in many emerging applications of communication technology, e.g., smart cities \cite{werbos2011computational} or internet of things \cite{tsai2014data}. Here, topics such as monitoring, fault prediction and scheduling are addressed with modern learning algorithms. 
%Since these applications often rely on in-network computation mechanisms, it is relevant to understand the trade-offs that are involved in designing such communication systems.
The use of machine learning methods in communications may provide information about individuals that affect their privacy.
Therefore, various privacy-preserving approaches to data analysis have been recently proposed (e.g., \cite{agrawal2000privacy}).
Machine learning methods are also widely applied to tackle security-related problems in communications, e.g., as part of defense mechanisms against spam attacks and viruses \cite{guzella2009review}.

The increasing convergence can be also observed in specific domains of communications such as image and video communication. While the direct approach to designing compression algorithms using autoencoders has provided very limited results compared to the state-of-the-art, the use of machine learning as an enhancing component for aspects like video encoding, bit allocation or other parts became a promising research direction \cite{zhang2015machine}. As most video signals are stored as compressed data, the topic of object recognition and tracking in the compressed domain is also of high relevance \cite{SriICASSP17}. Video streaming is another application which benefits from the use of learning algorithms \cite{liang2004real}.

Despite the successful use of machine learning methods in various communications applications, there are still many challenges and questions that need to be addressed.
For instance, the large size and high computational demands of modern machine learning algorithms prevent the large-scale use of these models in embedded devices. Also 5G networks call for novel machine learning-based approaches to radio resource management and network management approaches that can cope with uncertainties and incomplete channel and network state information.
Other problems concern reliability, privacy and security aspects of machine learning models. 

The following sections review the literature (Section \ref{sec:mlcomm}), present two applications (Section \ref{sec:application}) and discuss future research topics in machine learning and communications (Section \ref{sec:futuretopics}).
The paper concludes with a summary (Section \ref{sec:conclusion}).

%%%%%%%%%%%%%%%%%%%%%%%%%%%%%%%%%%%%%%%%%%%%%%%%%%%%%%%%%%%%%%%
\section{Machine Learning in Communications}
\label{sec:mlcomm}
\begin{figure*}[t]
\centering
\includegraphics[width=0.67\textwidth]{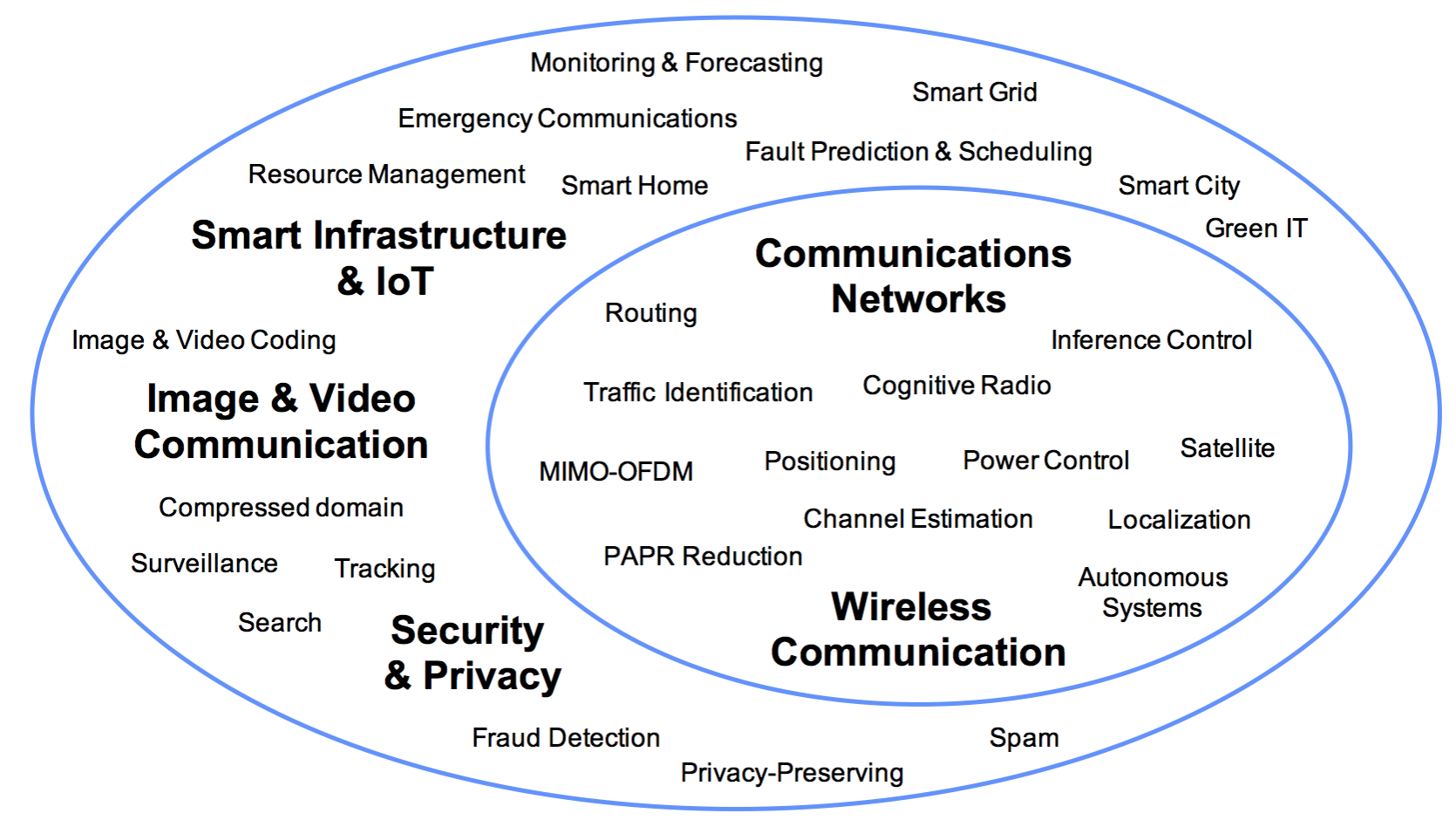}
\caption{Applications on machine learning in different areas of communications.}
\label{fig:overview}
\end{figure*}

This section reviews the use of machine learning algorithms in different application fields of communications (see Fig.\ \ref{fig:overview}).

%%%%%%
\subsection{Communication Networks}
Routing has a significant impact on the network's performance and is a well-studied topic in communications. Machine learning methods have been used to tackle different types of routing problems in the past, including shortest path routing, adaptive routing and multicasting routing.
The authors of \cite{boyan1994packet} proposed an algorithm for package routing in dynamically changing networks based on reinforcement learning. This algorithm learns a routing policy which balances between the route length and the possibility of congestion along the popular routes. 
Extensions on this idea have been proposed in \cite{kumar1997dual}.
Other researchers approached the routing problem with genetic algorithms \cite{munetomo1998migration}. Here alternative routes are created by crossover and mutation of the existing routes. 
Genetic algorithms have been also used for tackling the multicasting routing problem which emerges when data is send to multiple receivers through a communication network \cite{zhang1999orthogonal}. 
Also in mobile ad hoc networks the construction of multicast trees 
has been addressed using genetic algorithms.
Here additional objectives such as bounded end-to-end delay and energy efficiency are added to the optimization \cite{lu2013genetic}.

Several works (e.g., \cite{eswaradass2005neural}) have also used machine learning techniques for throughput or traffic prediction in communication networks. 
This is an important topic as with a dynamic throughput control and allocation one can fulfill the quality of service (QoS) requirements while efficiently utilizing the network resources.
For instance, the authors of \cite{liang2004real} applied neural networks for variable-bit-rate video traffic prediction in order to dynamically allocate throughput for real-time video applications. 
Traffic identification is another important topic for network operators as it helps them to manage their networks, to assure the QoS and to deploy security measures. 
Here, machine learning methods recognize statistical patterns in the traffic data by analyzing packet header and flow-level information. An excellent review of traffic classification with machine learning methods is \cite{nguyen2008survey}.

%%%%%%
\subsection{Wireless Communications}
To achieve a high efficiency at the desired QoS, it is essential in wireless systems to continuously adapt different parameters of MIMO-OFDM systems, in particular the link parameters, to the variations in the communication environment.
Various works (e.g., \cite{yun2010reinforcement}) tackle this parameter selection problem using machine learning.
Due to the dynamic nature of the wireless communication environment, there is also a strong need for adapting hardware parameters, e.g., to select a suitable set of transmit and receive antennas \cite{joung2016machine}.

The problem of reducing the peak-to-average power ratio (PAPR) is one of the key aspects in the design of OFDM-based wireless systems. 
Therefore, the problem has attracted much attention for many years. Application examples of machine learning to the PAPR reduction problem include neural networks \cite{jabrane2010reduction} and set-theoretic approaches \cite{cavalcante2009flexible} that are particularly suitable for online learning. Methods of machine learning and compressive sensing can also provide a key ingredient in enhancing the efficiency of OFDM channel estimation. For instance, the authors of \cite{cheng2016channel} address the problem by considering a neural network with known pilot signals at its input and the corresponding channel response at its output. Other works (e.g., \cite{sanchez2004svm}) turn their attention towards the problem of channel estimation in MIMO systems in the presence of nonlinearities. Learning-based approaches have been also applied for the estimation of mmWave channels \cite{mo2014channel}.

In order to enable an efficient and reliable opportunistic spectrum access, several approaches based on supervised, unsupervised, or reinforcement learning have been proposed in the literature. For instance, the study \cite{thilina2013machine} considers a cognitive radio system with cooperative spectrum sensing where multiple secondary users cooperate to obtain robust spectrum sensing results. Other approaches \cite{cavalcante2013distributed} apply distributed adaptive learning to tackle this problem.

Power control is a key mechanism for resource allocation in wireless systems. Machine learning has attracted some attention in the context of MIMO power control (e.g., \cite{mertikopoulos2016learning}). Various learning-based approaches (e.g., \cite{galindo2010distributed}) have also been proposed to tackle the inter-cell interference problem, which may have a detrimental impact on the performance of wireless users in mobile networks.
Furthermore, human supervision is still an indispensable element of current network management tools that are used to operate and manage mobile networks.
Much research effort has been spent in the last decade to fully automate the network management process and with it to realize the vision of self-organizing networks that operate without human intervention (see \cite{aliu2013survey}).

Information on the position of wireless devices is a key prerequisite for many applications. 
Machine learning methods have been used for localization \cite{zhang2010mobile} as well as navigation and positioning in car-to-car communication systems \cite{skog2009car}. 
%Furthermore, learning-based techniques have been used to process radar and sonar signals \cite{zhao2001support, barshan2000neural} and for signal pre-distortion in satellite communications \cite{goetz1996line}.

%%%%%%
\subsection{Security, Privacy \& Communications}
Machine learning methods play a pivotal role in tackling privacy- and security-related problems in communications.
For instance, they monitor various network activities and detect anomalies, i.e., events that deviate from the normal network behavior. Various machine learning methods have been applied for network anomaly detection in the past (see \cite{tsai2009intrusion}).
Other security applications are automatic spam filtering \cite{guzella2009review} and phishing attack detection \cite{basnet2008detection}.
Preserving data privacy is an important security aspect in communications, especially when sensitive data is involved. The design of machine learning algorithms that respect data privacy has recently gained increased attention.
The authors of \cite{agrawal2000privacy} demonstrated that it is possible to build a decision-tree classifier from corrupted data without significant loss in accuracy compared to the classifiers built with the original data, while at the same time it is not possible to accurately estimate the original values in the corrupted data records. This way one can hide private information from the algorithm, but still obtain accurate classification results. 

%%%%%%
\subsection{Smart Services, Smart Infrastructure \& IoT}
With the recent advances in communication technology the new field of ``smart'' applications attracted increased attention (e.g., smart homes, smart cities, smart grids, internet of things). Machine learning algorithms are often the core part of such applications.
For instance, the authors of \cite{ciabattoni2013design} used a neural network based prediction algorithm to forecast and manage the power production of a photovoltaic plant.
Other researchers applied similar techniques to traffic light control \cite{wiering2000multi} in smart cities or context aware computing in IoT \cite{perera2014context}.
Machine learning can also help detecting malicious events before they occur, e.g., in smart grid networks \cite{fadlullah2011early}.
Tasks such as prediction of a resource usage, estimation of task response times, data traffic monitoring and optimal scheduling have also been tackled with learning algorithms \cite{xu2008autonomic}.

%%%%%%
\subsection{Image \& Video Communications}
Machine learning methods have been used for various tasks in multimedia communication and processing (e.g., more than 200 applications of neural networks for images are summarized in \cite{egmont2002image}). Signal compression is one important field of application of these methods as it is part of almost every multimedia communication system. A survey on image compression methods with neural networks can be found in \cite{jiang1999image}. 
Tracking is another well-studied topic in machine learning which is also relevant in multimedia communication. A new generation of object tracking methods based on deep neural networks have been described in \cite{chen2016cnntracker}. Tracking algorithms which make use of the compressed video representation have also gained attention recently \cite{SriICASSP17}. 
%These methods are extremely fast and memory efficient as they only require to partially decode the video (i.e., extract motion vectors).
In multimedia applications such as video streaming the quality of the displayed video is of crucial importance. Different machine learning methods have been proposed to estimate the subjective quality of images perceived by a human \cite{BosICIP16, BosSMC16}.

%%%%%%%%%%%%%%%%%%%%%%%%%%%%%%%%%%%%%%%%%%%%%%%%%%%%%%%%%%%%%%%
\section{Exemplar Applications in Wireless Networking}
\label{sec:application}

The design and operation of wireless networks is a highly challenging task. On the road to the fifth generation of mobile networks (5G), researches and practitioners are in particular challenged by a multitude of conflicting requirements and promises as ever higher data-rates, lower-latency and lower energy consumption. 

The main cause of the problems and limitations in the context of 5G is the radio propagation channel. This so-called \emph{wireless channel} can strongly distort transmission signals in a manner that varies with frequency, time, space and other system parameters. The channel distortions are therefore of random nature and are notoriously difficult to estimate and predict. In addition, the wireless channel is a shared communication medium so that different wireless (communication) links must share the available communication resources. In modern mobile networks, this leads to interference between different mobile users, which in turn may have a detrimental impact on the network operation. As a result, the capacity of wireless links is of an ephemeral and highly dynamic nature, and it depends on global channel parameters such as path loss, path delay and carrier phase shifts, all of which vary with time, frequency and space.

Against this background, it is not surprising that the problem of reconstructing, tracking and predicting channel parameters play a prominent role in the design and operation of modern wireless networks such as 5G. Traditional approaches to this problem are usually based on the assumptions that 1) the wireless channel can be modeled with a sufficient accuracy and 2) a sufficient number of pilot-based channel measurements can be performed in real-time. However, the continuously increasing need for high spectral efficiency and the utilization of extremely high frequencies (above 6 GHz) makes these assumptions untenable in future networks. A potential solution will not be an adaptation or extension within an existing framework, but rather a paradigm shift is necessary to meet the requirements of 5G networks.  This in turn requires large strides both with respect to theoretical foundations and practical implementations. 

Modern wireless networks collect and process a huge amount of data and this data (including measurement data) can be used for tackling the mentioned problem of channel reconstruction, tracking and prediction. Therefore, in this context, a special attention has been attached to the development of new machine learning algorithms that are able to process spatially distributed data in \emph{real time} while efficiently using scarce wireless communication resources. This calls for the development of distributed algorithms that in addition must provide robust results, have good tracking (online) capabilities, and exhibit a relatively low complexity. Finally, they need to exploit context and side information such as spatial and temporal sparsity in the wireless channel. 

In the following subsection, we present one promising machine learning approach to the problem of reconstructing and tracking path loss maps in cellular networks. Subsection \ref{sec:application_compressedsensing} exemplifies the possibility of designing deep neural networks that exploit sparsity in the input data and amenable to real time implementation.

\subsection{Reconstruction of Radio Maps}
\label{sec: application_radiomaps}

\begin{figure}[htbp]
\centering
\includegraphics[width=0.45\textwidth]{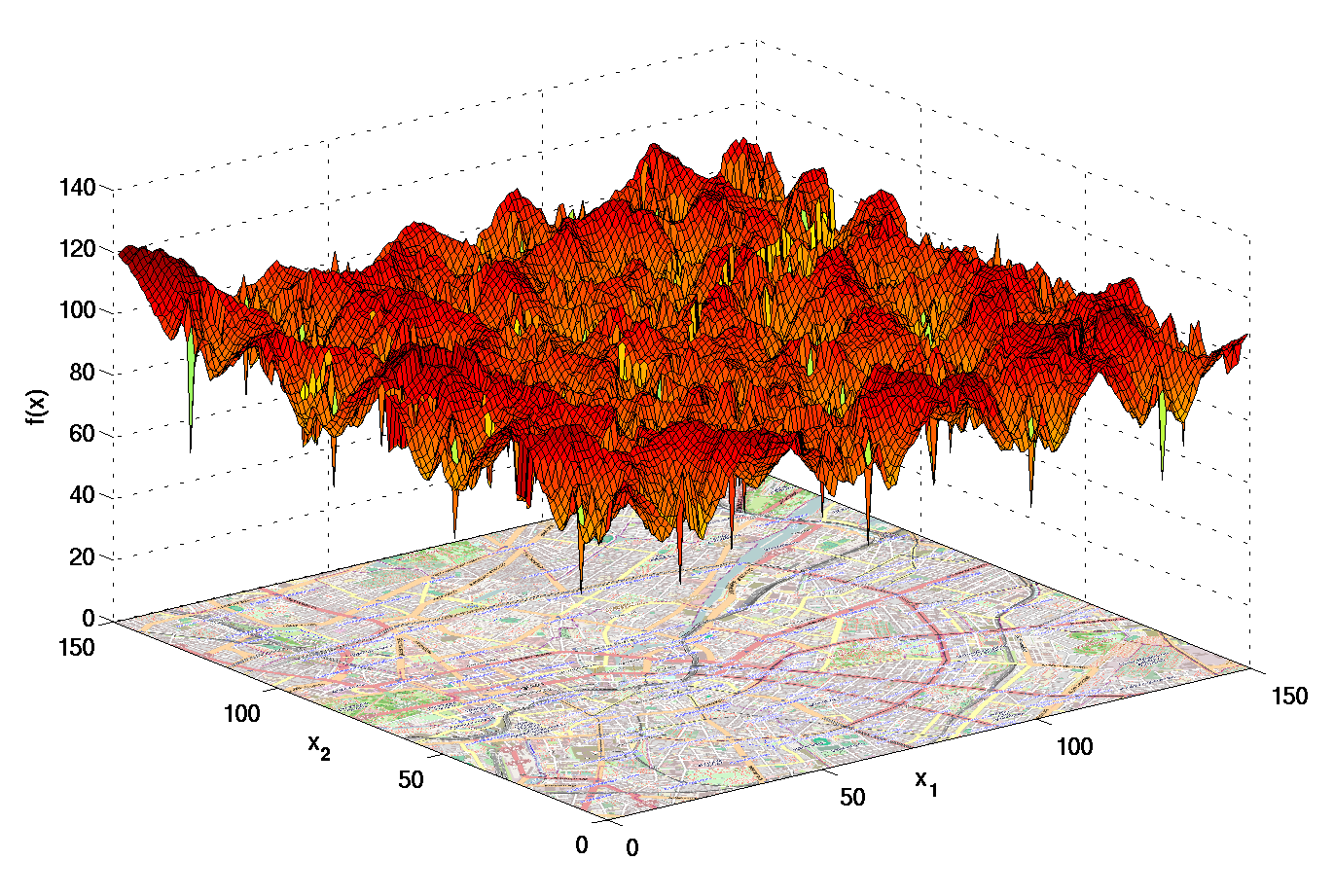}
\caption{An example of a path loss map for the downlink scenario with multiple base stations. The path loss map is a 2-dimensional function that assigns to a geographical position its path loss to the strongest base station.}
\label{fig:pathlossmap_multiplebs}
\end{figure}

We consider the downlink of a cellular network in which a number of base stations (transmitters) send the data to mobile users. While the users move over some geographical area covered by the network and send periodically their path loss measurements to the base stations, the problem is to reconstruct and track a (complete) path loss map in an online fashion from these measurements. The path loss determines the reduction in the power density of a transmission signal as it propagates from a base station to some geographic position. Note that for every geographical position, its path loss is defined with respect to the strongest base station, which is the base station with the smallest path loss. A radio map is then a function $f: \mathbb{R}^2\mapsto\mathbb{R}_{\geq 0}$ that assigns to every geographic position in a given area its path loss associated with the strongest base station. Fig.\ \ref{fig:pathlossmap_multiplebs} shows an example of a path loss map (the $2$-dimensional function over the geographic area) for the downlink of a cellular network. 
% \begin{figure}[htbp]
% \centering
% \includegraphics[width=0.45\textwidth]{figures/one_bs_3d_with_bs_1}
% \caption{Path loss map for a situation with a single base station}
% \label{fig:pathlossmap_onebs}
% \end{figure}

The general setting is as follows: Each base station collects path loss measurements sent by a subset of mobile users and it updates its estimate of the path loss map in an \emph{online} manner whenever a new measurement arrives. Measurements may contain errors since geographic location cannot be determined with arbitrary precision and measured path loss values can be erroneous. Finally, measurements are not uniformly distributed over a given geographic area so that more measurements may be available for some subareas than for others. \emph{The challenge is to reliably reconstruct the path loss map, including the path loss values for geographic positions for which no path loss measurements are available.} 

The problem was considered in \cite{kasparick2015kernel} where the authors propose using a multi-kernel approach based on adaptive projection methods. To be more precise, consider an arbitrary base station and let  $(x_n,y_n)\in\mathbb{R}^2\times\mathbb{R}$ be its measurement at time $n\in\mathbb{N}$, where $x_n\in\mathbb{R}^2$ is a sample (position measurement) at time $n$ and $y_n\in\mathbb{R}_{\geq 0}$ is the corresponding response (a noisy path loss measurement). An estimate $\hat{f}:\mathbb{R}^2\mapsto\mathbb{R}_{\geq 0}$ of the path loss map must be consistent with the available measurements. To this end, we require that $\forall_{n\in\mathbb{N}}\,|y_n-\hat{f}(x_n)|\leq\epsilon$ for some suitably chosen small $\epsilon>0$.  In \cite{kasparick2015kernel}, this requirement is met by projecting the estimate $\hat{f}$ on the hyperslabs given by 
$S_n=\{f\in\mathcal{H}: |y_n-\langle f,\kappa(x_n,\cdot)\rangle\},n\in\mathbb{N}$
where $\mathcal{H}$ is a reproducing kernel Hilbert space (RKHS) and $\kappa: \mathbb{R}^2\times\mathbb{R}^2\mapsto\mathbb{R}$ is the reproducing kernel for $\mathcal{H}$ so that $\langle f,\kappa(x_n,\cdot)\rangle=f(x_n)$ (the reproducing property). For lack of space, we refer the reader to \cite{kasparick2015kernel} for a rigorous definition of the concept of RKHS.  
 
Since $S_n$ is a closed convex set, the method of projection on convex sets (POCS) \cite{gubin67} provides the basis for the development of an iterative algorithm.  However, the POCS framework cannot be directly applied to our problem at hand because the number of measurements grows without bound as time evolves. Therefore, the authors of \cite{kasparick2015kernel} considered a different algorithmic approach that is a special case of the adaptive projected sub-gradient methods (APSM) developed in \cite{yamada04,renato09,renato12}. These methods open up the door to distributed implementation and real-time online processing via adaptive parallel projections on closed convex sets such as the hyperslabs. Moreover, they allow for incorporating context information in a systematic manner, while exhibiting relatively low-complexity and robustness against errors. For more details the reader is referred to \cite{yamada04,renato09,renato12,kasparick2015kernel}. 

\begin{figure}[htbp]
\centering
\includegraphics[width=0.45\textwidth]{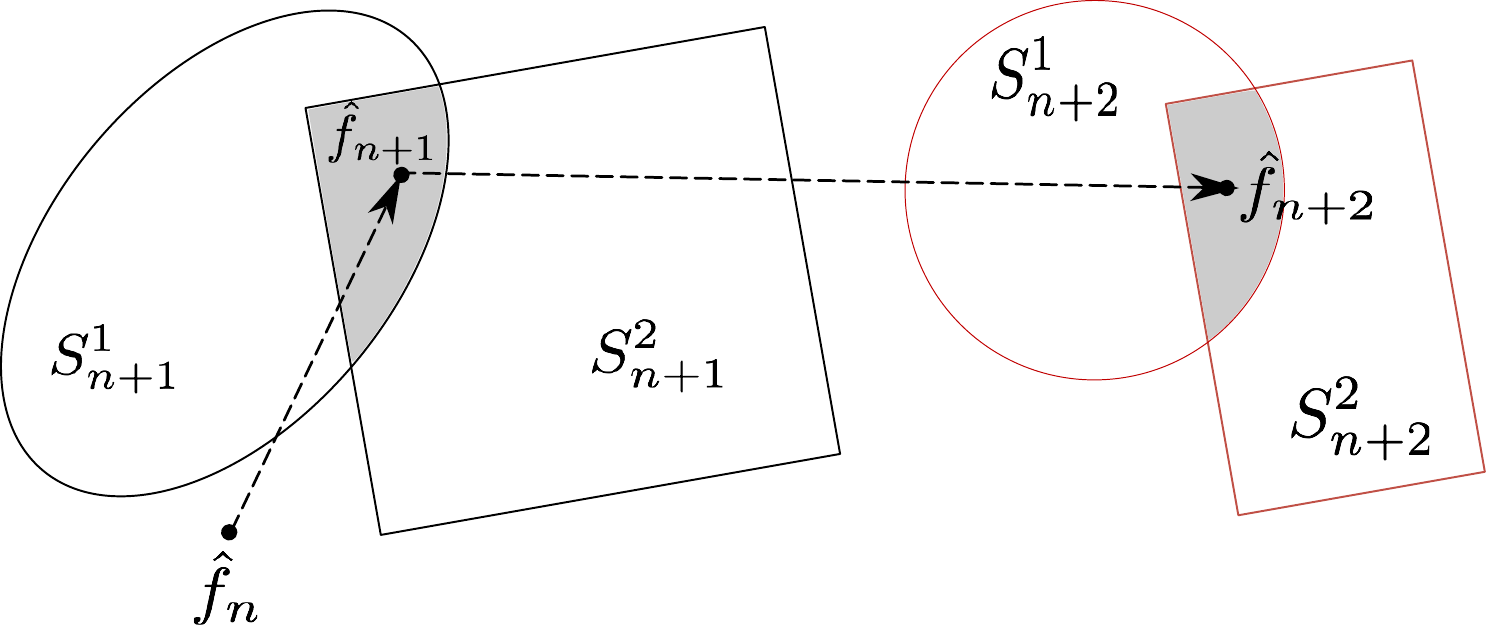}
\caption{Illustration of the APSM-based approach: Using parallel projection methods, the estimate $\hat{f}_n$ follows the intersections of the hyperslabs at times $n+1$ and $n+2$ to ensure consistency with new measurements and good tracking capabilities with online processing. At each time, there are two hyperslabs corresponding to two measurements.}
\label{fig:apsm}
\end{figure}

The main disadvantage of the APSM-based approach is the need for choosing appropriate kernel functions. In fact, in practical scenarios, different geographical positions require different kernel functions that in addition need to be adapted over time due to the dynamic nature of the wireless environment.  Since a real-time optimization of the kernel functions is an intricate task, inspired by the work \cite{yukawa12}, the authors of \cite{kasparick2015kernel} developed a multi-kernel approach that adapts kernel functions over time and space by choosing them from a large set of pre-defined kernel functions, while maintaining low-complexity and real-time capabilities. In the following, we briefly explain this approach. 

To this end, let $\{\kappa_m\}_{m=1}^M$ with $\kappa_m:\mathbb{R}^2\times\mathbb{R}^2\mapsto\mathbb{R}$ be a given set of some pre-defined kernel functions, where $M\gg 1$ is sufficiently large to include all relevant kernel functions. Since the number of measurements grows linearly with $n$, we take into account only the most relevant data which are contained in the dictionary $\{(x_i,y_i)\}_{n\in\mathcal{I}_n}$ where $\mathcal{I}_n\subseteq \{n,n-1,\ldots,1\}$ is the dictionary index set at time $n$. The cardinality $I_n=|\mathcal{I}_n|$ of the dictionary must be sufficiently small to satisfy the hardware limitations on memory size and processor speed. With these definitions, for an arbitrary time $n$, the estimate $\hat{f}_n(x)$ of the path loss at position $x$ can be written as a weighted sum of kernel functions: $\hat{f}_n(x)=\langle A_n, K_n\rangle=\mathrm{trace}(A_n^TK_n)$. Here $K_n=K_n(x)\in\mathbb{R}^{M\times I_n}$ is a given kernel matrix (evaluated at $x$) with $[K_n]_{i,m}=\kappa_m(x,x_i)$, and $A_n\in\mathbb{R}^{M\times I_n}$ is a parameter matrix that needs to be optimized. We point out that since the kernel matrix depends on the position $x\in\mathbb{R}^2$, the parameter matrix should be optimized for different geographical positions. 

The most obvious requirement on the parameter matrix $A$ is that it must chosen to fit the estimate to the measurements. This can be achieved by minimizing the distance (with some suitably chosen metric) of $A$ from the set $S_n=\{A\in\mathbb{R}^{M\times I_n}:|\langle A,K_n\rangle-y_n|\leq \epsilon\}$ for some sufficiently small $\epsilon>0$. Since $M$ is large, the problem is however computationally prohibitive for many applications in wireless networks. Therefore, the authors of \cite{kasparick2015kernel} extended the objective function by adding to the distance metric two regularization terms that impose some sparsity in $A$ when the new regularized objective function is minimized. As a result, the approach not only fits the estimate function to the measurements but also discards irrelevant data in the dictionary and reduces the impact of unsuitable kernels. 

The regularized objective function provides a basis for the development of new iterative algorithms in \cite{kasparick2015kernel} based on the forward-backward splitting methods and sparsity-based iterative weighting methods. The algorithms provide good tracking capabilities for the problem of reconstructing and tracking time-varying path loss maps.  For more details, we refer the reader to \cite{kasparick2015kernel}. 

\subsection{Deep Neural Networks for Sparse Recovery}
\label{sec:application_compressedsensing}

Recently, compressed sensing and deep learning have emerged as theoretical and practical toolsets to unleash full potential and approach fundamental theoretical bounds - whether it be for pilot decontamination in channel estimation, user identification, activity detection or PAPR reduction.

While in many cases researches are well aware of optimal solutions - e.g. in terms of optimization problems for channel estimation using minimal number of pilots - implementing these solution in embedded devices is considered infeasible due to unpredictable termination times and incalculable loss of early stopping. In this regard, a provisional solution aimed at large-scale measurement campaigns and utitlizing black-box data-driven machine learning techniques. While this approach fits well with many imaging problems, it was soon stripped of its enchantment for communication systems due to the necessity of measuring and pre-processing RF-signals under diverse sets of environmental conditions resulting in extremely large training times and disappointing performance gains. In addition, there is still no commonly accepted neural-network de-facto standard or baseline architecture for particular communication problems akin to AlexNet or GoogleNet in the imaging domains. 
% \begin{figure}[h]
%\centering
%\includegraphics[width=0.43\textwidth]{figures/csdl_net.eps}
%\caption{Application of deep learning for sparse recovery.}
%\label{fig:csdl_net}
%\end{figure}
One step to close this important gap was made in \cite{LiSt17} by using multidimensional Laplace transform techniques to design optimal neural networks for a particular sparse recovery problems revealing a very intriguing connection between commonly employed neural-networks comprising weights, threshold functions, recitified linear (ReLU) and rectified polynomial (ReP) activation functions and volume and centroid computation problems over sparsity inducing sets.
\begin{figure}[h]
\centering
\includegraphics[width=0.38\textwidth]{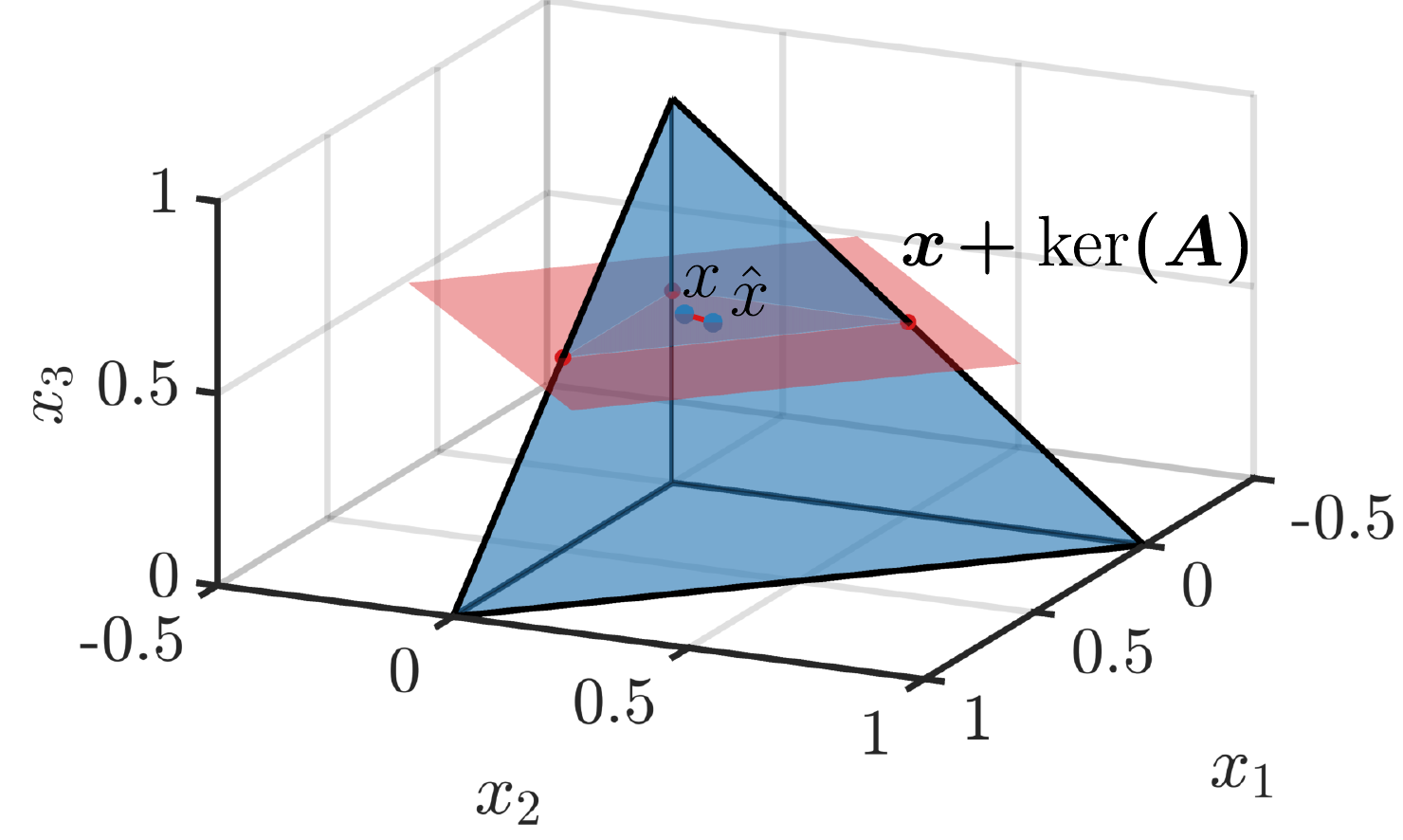}
\caption{Geometry of sparse recovery.}
\label{fig:csdl_geometry}
\end{figure}
We refer the reader to Fig.\ \ref{fig:csdl_geometry} for a geometric illustration of a small sparse recovery problem. Here, $\boldmath{x}$ is to be recovered from dimensionality reduced measurement $\boldmath{y}=\boldmath{A}\boldmath{x}$ given that $\boldmath{x}$ belongs to a particular sparsity inducing set (blue). Then, the neural-network of \cite{LiSt17} outputs the estimate $\hat{\boldmath{x}}$ that minimizes the expected error over the uncertainty cet (intersection between red and blue). Using such geometric ideas in the design of neural-networks allows for bypassing costly search over exponential candidate networks that consume large portions of available computing resources. Indeed, practitioners can still apply fine-tuning to reduce a possible model-mismatch and reduce reconstruction errors even further.

%%%%%%%%%%%%%%%%%%%%%%%%%%%%%%%%%%%%%%%%%%%%%%%%%%%%%%%%%%%%%%%
\section{Future Research Topics}
\label{sec:futuretopics}
This section discusses four future research topics in machine learning and communications.

%%%%%%
\subsection{Low Complexity Models}
State-of-the-art machine learning models such as deep neural networks are known to work excellently in practice.
However, since the training and execution of these models require extensive computational resources, they may not be applicable in communications systems with limited storage capabilities, computational power and energy resources, e.g., smartphones, embedded systems or IoT devices. 
Recent work addressed this problem and proposed techniques for reducing the complexity of deep neural networks.
For instance, the authors of \cite{han2015deep} demonstrated that the size of VGG-16, a popular deep neural network for image classification, can be reduced by over 95\% with no loss of accuracy. 
In communications applications such compression techniques can be used to store and transmit models efficiently.
Other authors (e.g., \cite{courbariaux2016binarized}) targeted the problem of weight binarization in deep neural networks.
This type of discretization can be useful, e.g., when adapting models to processor architectures which do not allow floating point operations.

Further research on these topics is of high importance as it can be expected that a large number of new applications would emerge, if the complexity of state-of-the-art models can be reduced to a level, which allows their use in computationally limited environments at minimal performance loss.

%%%%%%
\subsection{Standardized Formats for Machine Learning}
The standardization of algorithms and data formats is of high importance in communications, because it increases the reliability, interoperability and modularity of a system and it's respective components. With the increasing use of learning algorithms in communications applications, the need for standardized formats for machine learning is also rising.

For instance, standardized formats could be used to specify how to train, adapt, compress and exchange machine learning models in communications applications.
Furthermore, there could be standardized formats for the data and standards which determine how multiple machine learning models interact with each other.
Other formats could be specifically designed for ensuring that a model fulfills certain security or privacy requirements.

%%%%%%
\subsection{Security \& Privacy Mechanisms}
Machine learning models are often used in a black box manner in today's applications. This prevents the human expert from comprehending the reasoning of the algorithm and from validating its predictions. Although recent works \cite{BachPLOS15, MonArXiv17} proposed techniques for explaining the predictions of a machine learning model, further research on this topic is of high importance as the lack of transparency can be a large disadvantage in communications applications.

Moreover, it is well-known that deep neural networks can be easily fooled or may behave in an unexpected way when being confronted with data with different properties than the data used for training the model \cite{szegedy2013intriguing}. Thus, the establishment of mechanisms which increase the reliability of the model is a prerequisite for a large-scale use in communications applications. Such mechanisms can be implemented on different levels, e.g., be an integral part of the model, be integrated into the communication protocol or be part of a separate inspection process. 

Besides interpretability and security aspects, future research also needs to investigate how to effectively encrypt machine learning models and how to ensure data privacy during and after learning.

%%%%%%
\subsection{Radio Resource  and Network Management}
The end-to-end performance of mobile networks is strongly influenced by the choice of radio resource (e.g., beamforming and medium access control parameters) and network management (e.g., handover parameters, neighborhood lists, loads and power budgets) parameters. Moreover, some of the parameters must be continuously adapted on a relatively short time scale to time-varying radio propagation conditions and changing network topologies \cite{stanczak2009fundamentals}.

Current approaches are inadequate to cope with the growth of autonomous network elements in 5G small cell deployments based on mobile cloud RAN architectures. 
Therefore, 5G networks call for new model- and data-driven radio resource management and network management methods that are augmented by machine learning techniques for extracting knowledge from the system and gradual learning in the presence of inherent uncertainties and the lack of complete channel and network state information \cite{rondeau2007application}. 
The realization of these ideas in the context of 5G will require modifications of existing protocols and the development of new ones.

%%%%%%%%%%%%%%%%%%%%%%%%%%%%%%%%%%%%%%%%%%%%%%%%%%%%%%%%%%%%%%%
\section{Conclusion}
\label{sec:conclusion}
This paper discussed the increasing mutual influence of machine learning and communication technology.
Learning algorithms were not only shown to excel in traditional network management tasks such as routing, channel estimation or PAPR reduction, but also to be core part of 
many emerging application fields of communications technology, e.g., smart cities or internet of things.
The availability of large amounts of data and recent improvements in deep learning methodology will further foster the convergence of these two fields
and will offer new ways to optimize the whole communication pipeline in an end-to-end manner \cite{o2017deep}.

However, before resources-intensive models such as deep neural networks can be applied on a large scale in communications applications,
several practical challenges (e.g., complexity, security, privacy) need to be solved. Furthermore, more research is required on theoretical topics at the intersection of communications and machine learning, e.g., incremental learning, learning in nonstationary environments or learning with side information.

\bibliographystyle{abbrv}
\bibliography{IEEEabrv,bibliography}
\end{document}